\documentclass[iop]{emulateapj}

\usepackage[breaklinks=true]{hyperref} %% to avoid \citeads line fills
\hypersetup{colorlinks=true,urlcolor=blue,citecolor=blue,pdfborder= 0 0 0}

\slugcomment{}

\shorttitle{The lopsided distribution of satellite galaxies}
\shortauthors{Libeskind et al.}

\def\lsim{\mathrel{\lower0.6ex\hbox{$\buildrel {\textstyle <}
 \over {\scriptstyle \sim}$}}}
\def\gsim{\mathrel{\lower0.6ex\hbox{$\buildrel {\textstyle >}
 \over {\scriptstyle \sim}$}}}

\begin{document}


\title{The lopsided distribution of satellite galaxies}

%\author{Noam I. Libeskind\altaffilmark{1} and Quan Guo\altaffilmark{1}}

\author{Noam I. Libeskind and Quan Guo}
\affil{Leibniz-Institut f\"{u}r Astrophysik, Potsdam, An der Sternwarte 16, 14482 Potsdam, Germany}

\author{Elmo Tempel}
\affil{Tartu Observatory, Observatooriumi 1, 61602 T\~{o}ravere, Estonia}

\and

\author{Rodrigo Ibata}
\affil{Observatoire astronomique de Strasbourg, Universit\'{e} de Strasbourg,\\ CNRS, UMR 7550, 11 rue de l'Universit\'{e}, F-67000 Strasbourg, France}


%\altaffiltext{1}{Visiting Astronomer, Cerro Tololo Inter-American Observatory.
%CTIO is operated by AURA, Inc.\ under contract to the National Science
%\altaffiltext{2}{Society of Fellows, Harvard University.}
%\altaffiltext{3}{present address: Center for Astrophysics,
%    60 Garden Street, Cambridge, MA 02138}
%\altaffiltext{4}{Visiting Programmer, Space Telescope Science Institute}
%\altaffiltext{5}{Patron, Alonso's Bar and Grill}


\begin{abstract}
The distribution of smaller satellite galaxies around large central galaxies has attracted attention because peculiar spatial and kinematic configurations have been detected in some systems. A particularly striking example of such behavior is seen in the satellite system of the Andromeda galaxy, where around 80\% are on the nearside of that galaxy, facing the Milky Way. Motivated by this departure from anisotropy, we examined the spatial distribution of satellites around pairs of galaxies in the SDSS. By stacking tens of thousands of satellites around galaxy pairs we found that satellites tend to bulge towards the other central galaxy, preferably occupying the space between the pair, rather than being spherically or axis-symmetrically distributed around each host. The bulging is a function of the opening angle examined and is fairly strong -- there are up to $\sim$10\% more satellites in the space between the pair, than expected from uniform. Consequently, it is a statistically very strong signal, being inconsistent with a uniform distribution at the 5$\sigma$ level. The possibility that the observed signal is the result of the overlap of two haloes with extended satellite distributions, is ruled out by testing this hypothesis by performing the same tests on isolated galaxies (and their satellites) artificially placed at similar separations. These findings highlight the unrelaxed and interacting nature of galaxies in pairs.
\end{abstract}


%% subject header.  Objects should be in the appropriate "individual"
%% headers (e.g. quasars: individual, stars: individual, etc.) with the
%% additional provision that the total number of headers, including each
%% individual object, not exceed six.  The \objectname{} macro, and its
%% alias \object{}, is used to mark each object.  The macro takes the object
%% name as its primary argument.  This name will appear in the paper
%% and serve as the link's anchor in the electronic edition if the name
%% is recognized by the data centers.  The macro also takes an optional
%% argument in parentheses in cases where the data center identification
%% differs from what is to be printed in the paper.

\section{Introduction}
\label{introduction} 

The current paradigm of structure and galaxy formation, known as the $\Lambda$CDM model, asserts that the early universe started as a near homogenous density field, seeded by inflation with small density fluctuations. These small primordial density fluctuations constitute a Gaussian random field whose statistics were famously first computed by \cite{1986ApJ...304...15B}. The Gaussian nature of the over-densities implies that they are, to first order, modeled as spherically symmetric perturbations. However as structure forms due to gravitational instability, the initial anisotropy of the fluctuations become exacerbated such that by the present day, the dark matter haloes which form out of this process, are fairly aspherical. Haloes of mass $10^{12}M_{\odot}$ have axis-ratios of around 0.6 \citep[the axis ratio depends on many factors, including mass, redshift and cosmology, see][]{2006MNRAS.367.1781A}.

The shape of dark matter haloes can be probed by examining the location of satellite galaxies that inhabit them. It is a fairly well established empirical finding that satellite galaxies tend to not be isotropically distributed around their hosts. The ``flattening'' of satellite distributions is seen by both large surveys (i.e. in the 2dF or SDSS, \citealt{2004MNRAS.348.1236S, 2009MNRAS.395.1184S, 2005ApJ...628L.101B, 2007MNRAS.376L..43A, 2010ApJ...709.1321A}, see also \citealt{1997ApJ...478L..53Z}) and in the local Universe: satellites of the Milky Way \citep[e.g.][to name a few]{1976MNRAS.174..695L, 1994ApJ...431L..17M, 2005A&A...431..517K}, our Local Group partner M31 \citep{2013Natur.493...62I, 2013ApJ...766..120C, 2013MNRAS.435.2116P} and Centaurus~A \citep{2015ApJ...802L..25T} all exhibit a pronounced flattening.

Although the prevailing $\Lambda$CDM paradigm of structure formation does not obviously reproduce such extreme behavior \citep{2008ApJ...680..287M,2014MNRAS.442.2362P} a number of studies have been able to accommodate such set-ups by relying on either the triaxial nature of host haloes or large-scale filaments (e.g. \citealt{2005ApJ...629..219Z, 2005MNRAS.363..146L, 2009MNRAS.399..550L, 2013MNRAS.429.1502W, 2015ApJ...809...49B, 2015arXiv151101098S, 2015MNRAS.452.3838C, 2015MNRAS.453.3839P,2015ApJ...800...34G}). However debate still exists on whether the simulations are being appropriately compared with the observations. Other issues such as the stability of satellite planes and their formation mechanisms or how satellite galaxy planes relate to other well known problems of low mass objects (such as the ``missing satellite'' \citep{1999ApJ...522...82K} or the ``too big to fail'' \citep{2012MNRAS.422.1203B} problem) imply that it is clearly important to scrutinize the orientation of satellite galaxies around hosts. 

The anisotropic distribution of satellites is often couched in terms of 2-dimensional and implicitly axisymmetric systems. Yet in the case of M31, at least 21 out of 27 galaxies in the region covered by the PAndAS survey \citep{2013ApJ...766..120C} appear to be on the near side of M31, preferentially occupying the space in between the two main galaxies. Therefore not only are $\sim$50\% of M31's satellites confined to a thin, apparently spinning disc-of-satellites, but $\sim$80\% constitute a lopsided, highly azimuthally axis-asymmetric distribution. Quantifying such a lopsidedness around the Milky Way is complicated by both the zone of obscuration due to the Galactic disk, but also by the lack of complete sky coverage for ultra-faint dwarfs discovered by the SDSS. Despite the limitations and bias, if we consider just the ``classical'' 11 brightest Milky Way satellites, around 4 of them are on the side towards Andromeda. 

It is not clear if such a lopsidedness is commonplace among galaxies beyond the Local Group. This issue has motivated the work that follows. Do satellites of pairs beyond the Local Group show a statistically significant tendency to have a lopsided distribution with respect to the geometry of their hosts? A hint comes from work such as \cite{2008ApJ...675..146F}, which examined the tendency of satellites, albeit in $N$-body simulations, to be closer or further from principal directions of the matter distribution. Yet such studies implicitly gloss over the issue of lopsidedness, because they usually examine angles in the range $(0\degr,90\degr)$, instead of $(0\degr,180\degr)$. Regardless of this important subtlety, studies that find alignments between neighboring groups' shape, often consider objects orders of magnitude more massive than the Local Group, over larger scales in distance as well. Here we confine ourselves to galaxy pairs that roughly mimic the Local Group and quantify the tendency for satellites to fill the space between them.

The locations of satellites around central galaxies -- even in binary pairs like the Local Group -- is surely related to distribution of matter on larger scales, namely filamentary accretion along the so-called ``cosmic web'' \citep{1996Natur.380..603B}. On satellite scales, much work has focused on mainly two different types of anisotropic alignments: those with respect to the cosmic web \citep[i.e.][]{2011MNRAS.414.2029P,2014ApJ...791...15L,2015ApJ...799..212L,2015MNRAS.452.1052L,2015MNRAS.450.2727T}, and those with respect the central galaxy \citep[i.e.][]{2006MNRAS.369.1293Y,2008MNRAS.385.1511W,2014MNRAS.442.1363W}. Since in both of these cases the environment is often assumed to be the main driver of alignments, in this work we examine situations where the tidal field is dominated by a close partner. In order to isolate the effects of an overlapping satellite distribution, we compare our results with a carefully curated overlap sample.\\

\begin{figure*}
 \centering
 \includegraphics[width=40pc]{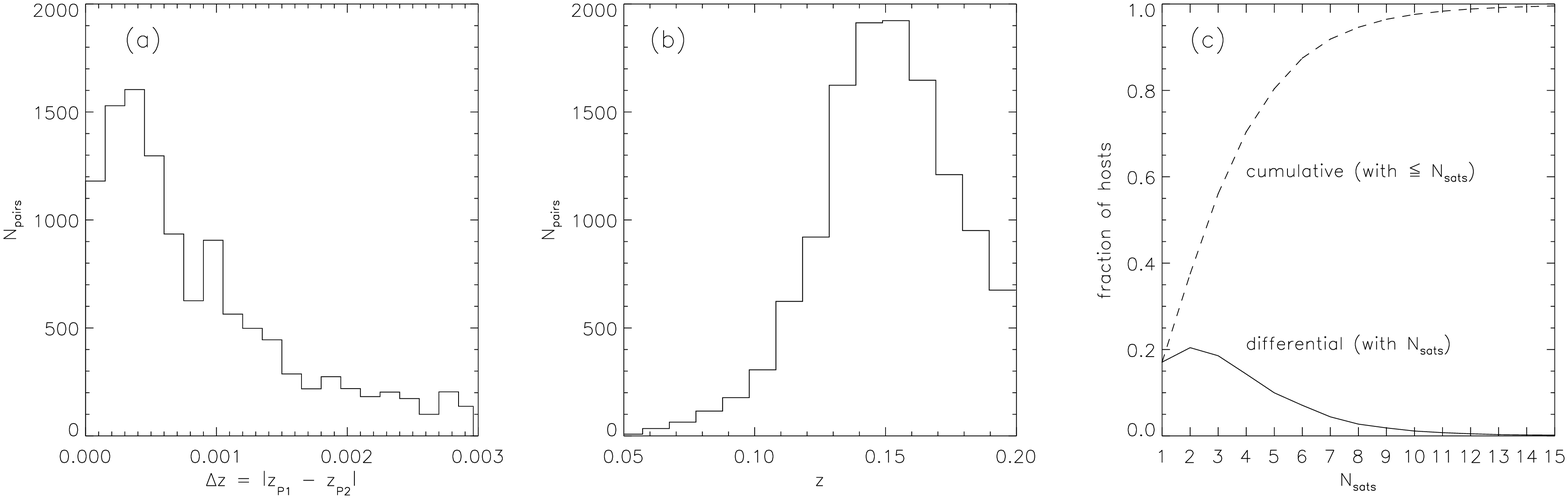}
  \caption{The distribution of redshift differences $z_{P1}-z_{P2} = \Delta z$ (left-hand panel) and of redshifts $z$ (middle panel) for the pairs of galaxies considered here (in the observational sample). By construction these are identical to the overlap sample. Right-hand panel: the number of the satellite galaxies around hosts is presented. The fraction of hosts that have $N_{\rm sats}$ is shown cumulatively (dashed) and differentially (solid).}
  \label{fig:zsh}
\end{figure*}

%%%%%%%%%%%%%%%%%%%%%
\section{Data and methods}
\label{section:methods}
In this section we explain how pairs of galaxies and their satellites are selected from the SDSS DR10 \citep{2000AJ....120.1579Y, 2014ApJS..211...17A} for our analysis. We select two samples -- an ``observational sample'' used to examine the hypothesis of this paper, and an ``overlap sample'' used to examine the effect of overlapping satellite distributions.

\subsection{Observational sample}
\label{sec:obs}

The first step to examine azimuthal asymmetry in satellite galaxy distributions is to identify pairs of galaxies in the sample. This is done by using the group finding method published by \cite{2014A&A...566A...1T}. Essentially, a Friends-of-Friends algorithm with variable linking length in the radial and transverse direction to account for redshift space distortions, is applied to the galaxy distribution. Only galaxies in the spectroscopic sample are considered for pair selection. We adopt cosmological parameters consistent with \cite{2014A&A...571A..16P} cosmology, the Hubble constant $H_0 = 67.8~\mathrm{km~s^{-1}Mpc^{-1}}$, the matter density $\Omega_\mathrm{m} = 0.308$, and the dark energy density $\Omega_\Lambda = 0.692$. Groups of only two galaxies are retained for further analysis, as these constitute a sample of galaxy pairs. With this method we obtain a set of 48\,917 two-galaxy pairs, which we subject to further cuts, outlined below. The preparation and details of the spectroscopic galaxy sample are described in \citet{2014A&A...566A...1T}.

We select pairs where each member has a magnitude in the $r$-band between ($-23.5< M_{r}^{0.0}<-21.5$) as our fiducial magnitude cut (this cut is examined in more detail in section 3.2.1). This magnitude range is motivated by estimates of the MW and M31's $r$-band magnitude from \cite{2011ApJ...733...62L}, which although not-identical (-21.2) is not too far off either. The magnitude range is purposefully kept wide in the interest of not prejudging our sample. Such that only binary systems where the two galaxies have similar magnitudes are considered, a further cut is performed and only pairs that are within a magnitude difference of each other are retained.

In order to avoid close pairs of galaxies, the two primaries are required to be separated by $0.5<d_{\rm sep}/\mathrm{Mpc} < 1.0$ (in projection). The mean redshift of the two primaries is taken as the distance to the pair and it is at this redshift that distances are computed. Changing $d_{\rm sep}$ has little effect since it is difficult to get groups of just two galaxies when $d_{\rm sep}$ is increased (few pairs have larger values of $d_{\rm sep}$). The value of $d_{\rm sep}$  is considered in more detail in section 3.2.4.

Once these cuts have been applied to the host, a (projected) search radius within which satellites are retained, is examined. The fiducial search radius is set to $r_{\rm search}=250$~kpc, roughly equal to the virial radius of a Milky Way mass galaxy halo. The search radius is examined in more detail in section 3.2.3.

The potential satellite galaxies are searched from the photometric catalogue of the SDSS. Since galaxy pairs are selected from the spectroscopic redshift catalogue and by construction they do not have other spectroscopic galaxy nearby, the potential satellite galaxies do not have measured (spectroscopic) redshifts. Because of this, the selection is done only in projection, hence, not all of these will be satellites in the strict sense (namely within a virial radius). Indeed many have poorly constrained photometric redshifts that are inconsistent with their alleged hosts. However, any interlopers are expected to be homogeneously distributed and therefore cannot dominate the signal. Therefore any signal obtained ignoring the photometric redshifts of the potential satellites should be seen as a lower limit.

In sum, there are four criteria which determine which hosts and satellites constitute the fiducial sample: 
\begin{itemize}
\item host magnitude: $-21.5<M_{r}<-23.5$;
\item host separation $0.5 < d_\mathrm{sep}/{\rm Mpc}<1.0$;
\item magnitude difference of pair $<$1 mag
\item satellites within a (projected) search radius: $r_{\rm search}=250$~kpc;
\item no consideration of satellite photometric redshift.
\end{itemize}
In Section~\ref{sec:sys}, a section devoted to systematic effects, we examine how the signal changes if these fiducial values are altered.

Our final fiducial sample consists 12\,210 pairs of galaxies which among them have 46\,043 potential satellites. In Fig.~\ref{fig:zsh}(a) we show the redshift difference of the pairs included in the observational sample (namely $\Delta z=|z_{\rm  P1} - z_{\rm P2} |$ where $z_{\rm P1}$ and $z_{\rm P2}$ are the redshifts of the brighter (P1) and fainter (P2) primary). These pairs tend to be close in redshift (by construction) with redshift differences of $\Delta z< 0.003$. Fig.~\ref{fig:zsh}(b) shows the redshift distribution of the mean redshift of the pair, while Fig.~\ref{fig:zsh}(c) shows how many satellites are found per host. The median satellite number per host is three, while the mode of the distribution is two satellites.

\begin{figure}
 \centering
 \includegraphics[width=\columnwidth]{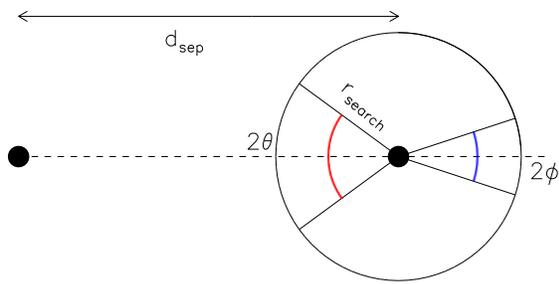}
  \caption{A schematic diagram indicating the geometry of galaxy pairs in the plane of the sky. A pair of primary galaxies (black points) are identified according to the algorithm defined in \cite{2014A&A...566A...1T} and described in Sect.~\ref{section:methods}. Satellite galaxies within the projected search radius $r_{\rm search}$ are retained as satellites. The line connecting the two primaries, $d_{\rm sep}$, is used to compute the angles $(\theta, \phi)$, which face towards or away from (respectively) the other primary.}
  \label{fig:schema}
 \end{figure}

\subsection{Overlap sample}
\label{sec:overlap}
An immediate area of concern regarding the effect considered in this paper, is contamination due to extended overlapping satellite distributions. Suppose two galaxies have accompanying satellite systems that extend well beyond the 250~kpc considered here. Consider placing these two galactic systems at the separations considered here, namely between 500~kpc$<d_{\rm sep}<$1000~kpc. In such a situation one would expect that the projected overlap of the satellite systems of these galaxies alone would result in a similar signal as the bulging signature we wish to isolate\footnote{We note that such an overlapping signal should have a maximum exactly in between the two primaries. If the luminosity of the primaries (or radial concentration of the satellites), is different, then the maximum overlapping signal should be slightly shifted toward the fainter primary.}. In order to control for this overlap effect we construct an ``overlap sample''. In practice we wish to replace each pair member from our observational sample with an isolated ``doppelg\"anger'' galaxy (and satellites).

Constructing a sample of overlapping pairs begins with a list of isolated primaries drawn from the spectroscopic sample of the SDSS. ``Isolated'' galaxies are defined as those FOF groups \citep[as described above e.g. see][]{2014A&A...566A...1T} composed of just one single galaxy. All satellite galaxies, within a projected distance of 2.5~Mpc of each isolated primary are identified from the photometric survey.

For each pair in the observational sample, we select 5 pieces of data: the redshifts ($z_{\rm P1}$ and $z_{\rm P2}$), the magnitudes ($M_{r,P1}$ and $M_{r,P2}$) and the separation ($d_{\rm sep}$) of the two primaries. For each primary (P$i$, where $i\in\{1,2\}$) in each pair in the observational sample, we wish to find an isolated system whose redshift ($z_{{\rm D}i}$, D for ``Doppelg\"anger'') and magnitude  ($M_{r,{\rm D}i}$)  are as close as possible to $z_{{\rm P}i}$ and $M_{r,{\rm P}i}$. We specifically force $|z_{{\rm P}i}-z_{{\rm D}i}|=\Delta z<0.001$ and  $|M_{r,{\rm P}i}-M_{r,{\rm D}i}|=\Delta M_{r}<0.05$, where $i\in\{1,2\}$ represent the two members of the pair. 

Once two isolated primaries have been identified as the pair's doppelg\"angers, these galaxies (and all their satellites within 2.5~Mpc) are given a projected separation equal to $d_{\rm sep}$. Note that the sense of orientation of their two satellite systems is completely arbitrary. Therefore in order to sample just the effect of extended radial distributions, the angular position of each satellite is randomized 100 times while keeping its radius with respect to its host fixed. In this way, the effect of the radial distribution of each pair member's satellite system and the strength of any overlap effect can be quantified, isolated and subtracted from the observed signal. Note that we could have randomized the angular position of the satellites an infinite number of times, since the overlap sample just measures the deviation from uniform signal. In principle, if we measure the statistical significance for the observed sample, we should compare it with the overlap sample, not with the uniform sample. In practice it is equivalent to drawing the error bars around the overlap sample's curve and not around the uniform signal (ie unity). From a practical point of view, this does not change our results nor conclusions.

The same cuts are then applied to the overlap sample, namely only those satellites that fall within 250~kpc of each primary are considered. This is now a combination of each primary's ``own'' satellites and some contaminants from the other primary system. Regardless of which host the galaxies originally belonged to, the angles ($\theta,\phi$) are then computed for all satellites within 250~kpc of each host. Note that by construction the isolated primaries used to build the overlap sample have the same magnitude, redshift, and separation distribution as the observational sample. 

The ``true'' signal is thus the observed signal minus the overlap signal. Although in many of our plots we show all three (observed, overlap and true signals), we emphasize the biases caused by overlapping satellite systems.

\subsection{Naming convention}
Two data sets are used in the paper: the observational (section \ref{sec:obs}) and overlap (section \ref{sec:overlap}) samples. We adopt the following convention (see Fig.~\ref{fig:schema}). An opening angle facing the other primary is denoted $\theta$, while an opening angle facing directly {\it away} from the other primary is denoted $\phi$. The angles are measured on the plane of the sky. These angles are centered on and hence symmetric about the line connecting the two primaries. Thus, for example, $\theta=90\degr$ corresponds to the hemisphere facing the other primary while $\phi=90\degr$ is the hemisphere facing away from the other primary. In what follows we will simply count the number of galaxies within $(\theta,\phi)$ for the observational and overlap samples, and compare these to that expected from a uniform distribution.

%%%%%%%%%%%
\section{Results}

\begin{figure} 
 \centering
 \includegraphics[width=\columnwidth]{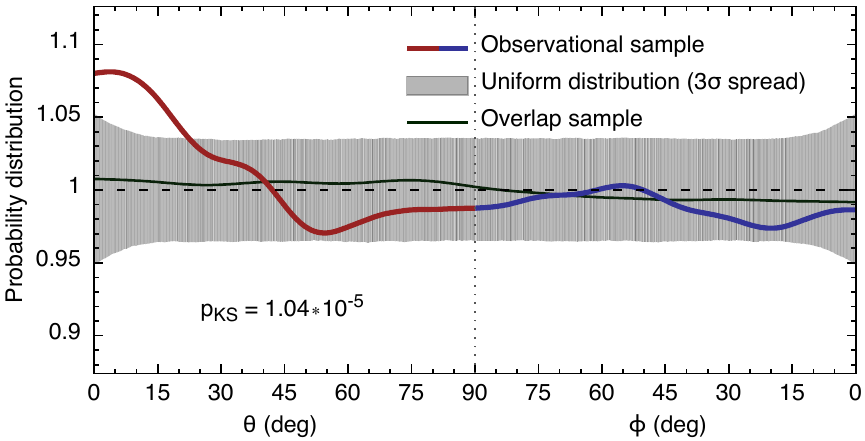}
  \caption{The probability distribution of the angles formed between the position vector of each satellite and the line connecting the two primaries. The red curve represents $\theta$ (the angle facing the other primary, see Fig.~\ref{fig:schema}) while the blue curve represents the distribution of $\phi$ (the angle facing away). The grey band shows the 3$\sigma$ region expected from a uniform distribution of the same size. The excess above this is seen at low $\theta$, and the fact that $\phi$ always remains well within 3$\sigma$ is the signature of a lopsided distribution. The Kolmogorov-Smirnov probability that the distribution is drawn from a random one is given as $p_{\mathrm{KS}}\sim 10^{-5}$ and rules out the null hypothesis at $\sim 5\sigma$ level. The distribution of angles from an artificially constructed overlapping signal  (see Section~\ref{sec:overlap}) is shown in green and constitutes a minor effect.}
  \label{fig:distr}
 \end{figure}

We quantify the anisotropic distribution of satellites by counting how many satellites are within a given opening angle $(\theta,\phi)$. The statistical significance of any anisotropy measured at each opening angle can be gauged via a simple Monte-Carlo test. We perform 100\,000 trials where each satellite galaxy's radial distance is kept fixed but its angular position around its host, is randomized. For each of these 100\,000 trials we count the number of (randomized) satellites within a given angle. This provides the expected deviation from a uniform distribution and allows for a statistical quantification of any measured signal. In what follows, these Monte-Carlo tests are what is used to statistically quantify the measured signals.

The distribution of angles made between the position vector of each satellite and the line connecting the two primaries is shown in Fig.~\ref{fig:distr}. In red we show $\theta$, the angle measured facing the other primary while in blue we show $\phi$, the angle measured facing away from the other primary (see Fig.~\ref{fig:schema}). The grey band represents the $3\sigma$ spread one expects from a uniform distribution of ($\theta,\phi$). The distributions are plotted on a single $x$-axis to highlight that because of the nature of our geometry ($\theta,\phi$) will never overlap. Since $\theta$ shows a strong excess at low values (say below $\sim 40\degr$), while $\phi$ is flat and well within that expected from a random, uniform distribution, we conclude a strong lopsidedness: {\it satellite galaxies tend to preferentially inhabit the regions in between galaxy pairs}. There are up to 8\% more satellites than expected from a uniform distribution at low values of $\theta$. A similar but significantly weaker effect is seen in the overlap sample (plotted in green) where the probability distributions always lie well within the $3\sigma$ interval expected from a uniform distribution, at around 1\%. Therefore we conclude that the lopsided effect seen in the observational sample is not the result of an overlapping satellite distribution. A Kolmogorov-Smirnov (KS) test can be performed on the full set of angles to quantify the likelihood that these are consistent with being derived from a uniform distribution and is indicated within Fig.~\ref{fig:distr} as $p_{\rm KS}=1.04\times10^{-5}$. Therefore, for the observational sample, {\it this hypothesis is rejected at the 5$\sigma$ level}. The overlap sample is fully consistent with uniform and the same hypothesis is not rejected by the KS-test.

\begin{figure}
 \includegraphics[width=20pc]{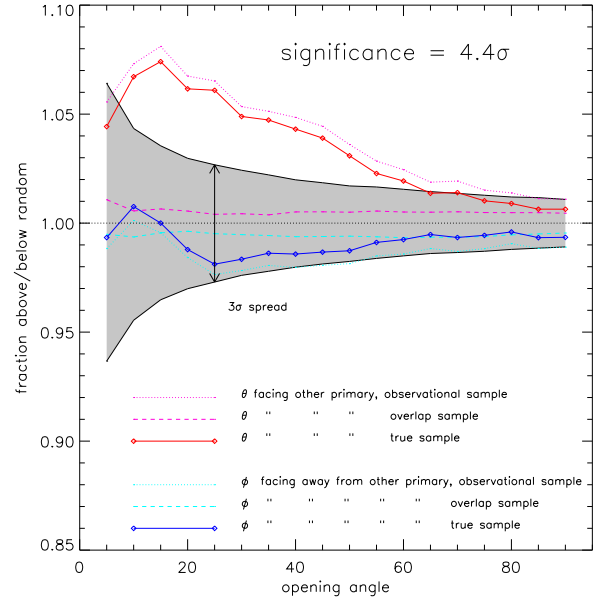}
  \caption{The over-abundance of satellites within a cone facing (red curve) or facing away (blue curve) from the other primary is shown by simply dividing the measured excess by that expected from a uniform distribution. Clearly, angles facing the other primary always have more satellites than angles facing away -- even at 90\degr, the hemisphere facing the other pair has more galaxies than the hemisphere facing away. The robustness of this claim is made by counting the number of galaxies within the same angle given uniform randomized positions. We randomize the position angle (with respect to the primary) of each satellite 100\,000 times and examine the spread in expected over- or under-abundances. The 3$\sigma$ spread is shown by the grey band. Within around 20$\degr$--40$\degr$ the bulging effect is a 5$\sigma$ deviation from random. The dashed lines show the effect expected if satellite distributions simply overlap. For each pair in the observational sample two isolated galaxies (that have similar redshifts and magnitudes) are placed at the same separation. The angles, with respect to the line connecting the two galaxies, of all the ``satellites'' within 2.5~Mpc of each host are then randomized, while keeping the projected distance of the satellite to the host fixed. Such an overlap effect constitutes a $\sim1\%$ effect, is always well within $3\sigma$ and thus cannot explain the bulging seen in the solid curves.}
  \label{fig:cone}
 \end{figure}

The bulging effect can be quantified by asking a complimentary question: ``how many more satellites are seen, {\it within} a given opening angle than expected from a uniform distribution?'', essentially a cumulative version of Fig.~\ref{fig:distr}. This is shown by the magenta (for $\theta$) and cyan (for $\phi$) curves in Fig.~\ref{fig:cone} for the observed (dotted) and for the overlap (dashed) samples. The so-called ``true'' signal is simply the observed distribution divided by the overlap sample and is shown by the solid blue and red curves. The true signal in Fig.~\ref{fig:cone} shows that within the strongest, most over-abundant region, $\sim 30\degr$, bulging approaches an 8\% effect. As opening angles increase, the bulging effect decreases, becoming just consistent with uniform (at the $\sim2\sigma$ level) at $180\degr$, namely within the hemisphere.

The strength of the lopsided signal can be ascertained by computing, in units of $\sigma$, the average deviation from uniform of a given signal. For the fiducial sample shown in  Fig.~\ref{fig:cone}, the significance of the observed sample is more than $4\sigma$.

The relative abundance of satellites with respect to the angle facing {\it away} from the other primary (namely $\phi$) is roughly constant below unity but within $\sim3\sigma$ for the observed sample. This means that there is aways a deficit of satellite galaxies on the far side of the pair, but this deficit is well within 3$\sigma$ of what one would expect from uniform distributions of the same size. Any deficit is therefore not statistically significant and is consistent with a uniform distribution. Note that within all angles facing towards and away from the other primary, the overlap sample shows complete consistency with uniform distributions and is thus not sufficient to explain the measured signal. The overlap signal represents a departure from uniformity at the 1.09$\sigma$ level, while the observational signal for some angles ($\sim 20\degr$) is over a 5$\sigma$ effect.

When the effect of overlapping satellite distributions is subtracted from the observed signal, the final true signal is on average 4.4$\sigma$ away from uniform. Although the true signal ``only'' reaches a maximum of an 8\% increase over uniformity, this 8\% is statistically significant and robust. Furthermore, it is also a lower limit, because most of the ``satellite galaxies'' are interlopers.

\begin{figure*}
 \centering
 \includegraphics[width=0.95\columnwidth]{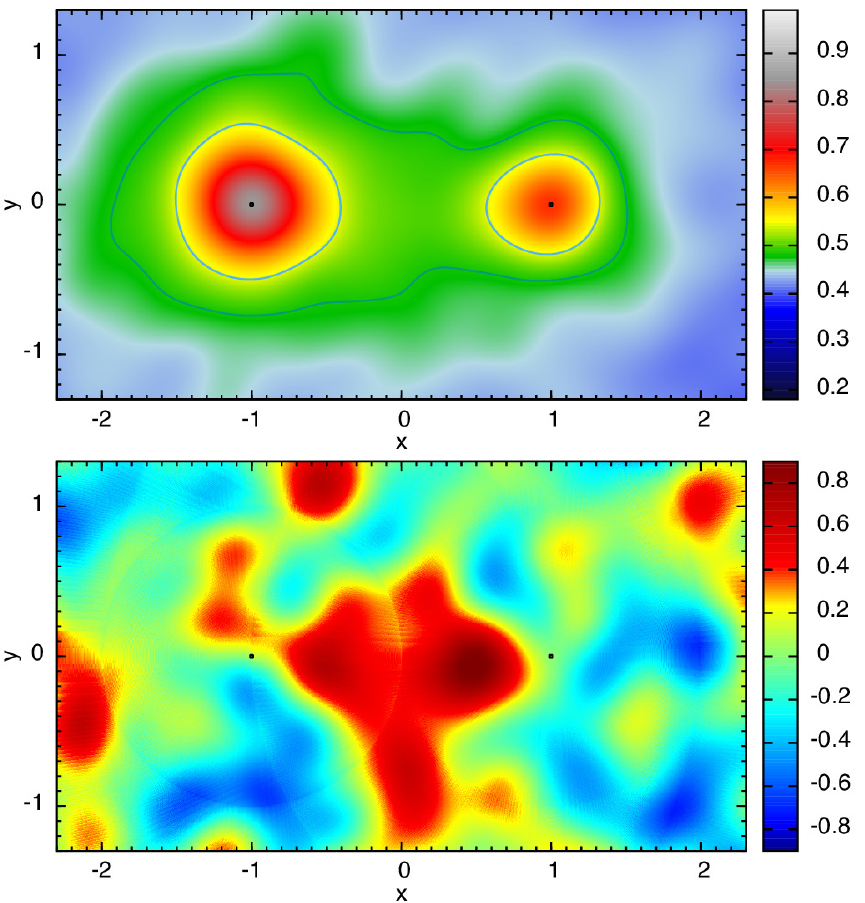}
 \includegraphics[width=0.95\columnwidth]{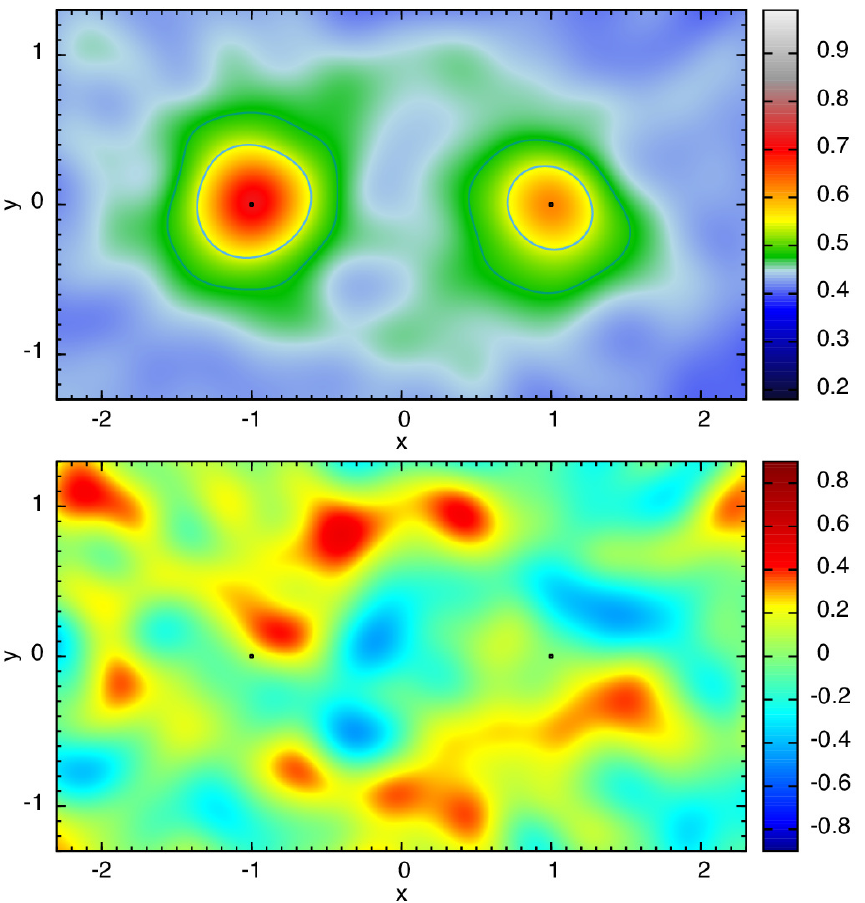}
  \caption{The bulging nature of satellite galaxies can be visualized in a 2-dimensional ``sky plot'' which mimics the view of these systems from the SDSS. Each primary is assigned a position (the bright one is placed at $x=-1$, and the fainter one at $x=1$) and the number density of satellites is simply contoured with red being large density and blue is low density (top panel). The color bar shows the number of satellites per arbitrary unit area. In the bottom panel the spherically averaged radial distribution computed while masking out the other primary and its satellites, is subtracted from each primary to highlight the deviation from spherical symmetry and anisotropic morphology of the satellite population. The color bars in the lower panels are arbitrary representations of where more satellites are found with respect to the spherically averaged distributions. {\it Left}: observed sample; {\it right}: overlap sample with the same number of objects as in the observed sample. }
  \label{fig:vis}
 \end{figure*}

%%%%%%%%%%%%%%%
The situation can be visualized by stacking all satellites of the galaxy pairs and examining the projected 2D number density of satellites, shown in the top left panel of Fig.~\ref{fig:vis}. Here, the brighter primary is given a position at $(x,y)=(-1,0)$ and the fainter primary is placed at $(1,0)$. The (projected) number density of satellites is then (arbitrarily) contoured and displays a bar-bell morphology with a clear bulging of satellite galaxies towards each other. The number density of satellites in Fig.~\ref{fig:vis} is the combination of foreground/background galaxies and true satellites. The foreground/background density can be estimated from the outer parts of the figure. While counting a satellites around host primaries and subtracting the foreground/background density, we are left with a true number of satellites. Altogether around 15\% of all galaxies are true satellites of the central galaxies, most of the galaxies are foreground/background objects.

The averaged radial distribution of satellites (masking out the other primary and satellites belonging to it) can be calculated for each host, directly from the top left panel of Fig.~\ref{fig:vis}. If this radial profile is subtracted from the satellite density map, we are left with a morphological impression of the anisotropic spatial distribution of the satellite population, shown in the bottom left panel of Fig.~\ref{fig:vis}. Here the over-density of satellites between the two primaries is clearly visible. In order to compare with the overlap sample, the same plots are shown on the upper and lower right of Fig.~\ref{fig:vis}. The morphologies are markedly different: the overlap sample shows no significant contaminatory signal from overlapping galaxies. Moreover, the morphology of the lopsided signal is different than expected from an overlapping satellite distribution.

 \subsection{Filamentary environment}
	One possible interpretation of our findings is that the galaxy pairs, which can be separated by $\Delta z \sim 10^{-3}\approx300~{\rm km/s}\approx 4~$Mpc, are aligned along the line-of-sight with a filament. In this interpretation, the lopsidedness results from the filament's potential itself, rather from the binary interaction of the galaxy pair. In order to test this hypothesis we identified which of our galaxy pairs were located in filaments identified with the Bisous method \citep[see][]{2014MNRAS.438.3465T,2016arXiv160308957T}. The Bisous method is a marked point process used to identify curvilinear patterns (such as filaments) in the galaxy distribution. It has been employed to study, among other things, satellites and galaxy alignments in the SDSS before \citep{2013ApJ...775L..42T,2015MNRAS.450.2727T,2015A&A...576L...5T,2016MNRAS.457..695P} and tested against other methods for identifying filaments \citep{2014MNRAS.437L..11T,2015MNRAS.453L.108L}. In general it is accepted as a reliable way of identifying such structures in the community.

\begin{figure*}
 \centering
 \includegraphics[width=2.0\columnwidth]{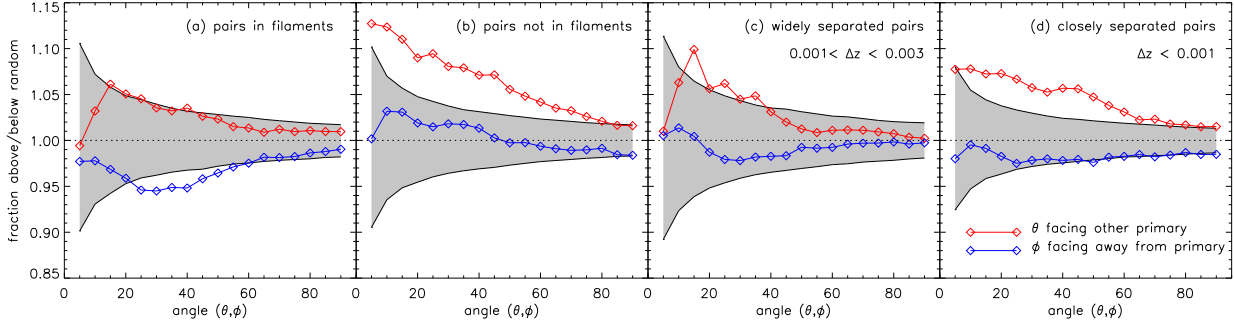}
  \caption{Examining the filament hypothesis. After identifying pairs that are in Bisous filaments \citep[see][]{2014MNRAS.438.3465T}, we compute the lopsided signal for pairs in such filaments (a) and pairs not in Bisous filaments (b). Pairs in filaments do not show a statistically significant signal. We also examine how the pair's separation (measured as the redshift difference of the two members $\Delta{\rm z}$) affects the signal, with widely (c, $0.001<\Delta{\rm z} < 0.003$) and closely (d, $\Delta{\rm z}<0.001$) separated pairs, under the assumption that if the signal was dominated by filaments, the signal would be stronger for widely separated pairs. The signal is roughly the same in both cases, although in the former (c), the sample size is smaller, so the statistical significance of any signal is weaker.}
  \label{fig:filaments}
 \end{figure*}

When applied to the sample here, galaxy pairs are ascribed a simple ``within'' or ``not in'' filament status in a similar way as defined in \citet{2015ApJ...800..112G}. The lopsidedness signal for these two subsamples is shown in Fig.~\ref{fig:filaments} (a and b). Pairs in filaments (Fig.~\ref{fig:filaments}a) show a weaker signal than those not found in filaments (Fig.~\ref{fig:filaments}b), although we note that the sample of the former is smaller and thus any statistical statement is weaker. The most straight forward interpretation of this result is that the lopsided signal is stronger for more isolated pairs, and that when pairs are found within the more dynamically active environment of filaments, the lopsided signal can be weakened. We note that we also examined the strength of the lopsided signal for in-filament pairs depending on whether the filament was oriented along or perpendicular to the line of sight (not shown). In this case when the filament is along the line of sight the signal is weaker, when its perpendicular the signal is stronger, further supporting this hypothesis. Again however, such statements are statistically weak, given the reduced sample size.

Another way of testing the filament hypothesis is to divide the sample by $\Delta z$ into ``closely separated'' pairs ($\Delta z< 0.001$) and ``widely separated'' pairs ($0.001 < \Delta z <0.003$). In Fig.~\ref{fig:filaments} (c and d) we show that the lopsided signal is driven by pairs that are close in line-of-sight separation. Further separated pairs show a weaker, although not altogether absent, lopsidedness in their satellite distribution. That is, our results do not support the filament hypothesis on filament scales of more than $\sim 4$ Mpc (which corresponds roughly to $\Delta z= 0.001$). 
 
Our method for finding filaments in the galaxy distribution - the Bisous model - could also be mis-identifying (or simply missing) filaments along the line of sight. In other words there could be filaments causing the effect reported here that we simply do not identify. Although our tests can not rule out this hypothesis conclusively, the evidence we have presented in Fig.~\ref{fig:filaments} is inconsistent with it.

\subsection{Systematic effects}
\label{sec:sys}
In this section we examine, systematically, how our choice of fiducial parameters may be driving the lopsided effect reported here.

\subsubsection{Host magnitude}
\begin{figure*}
 \centering
 \includegraphics[width=1.75\columnwidth]{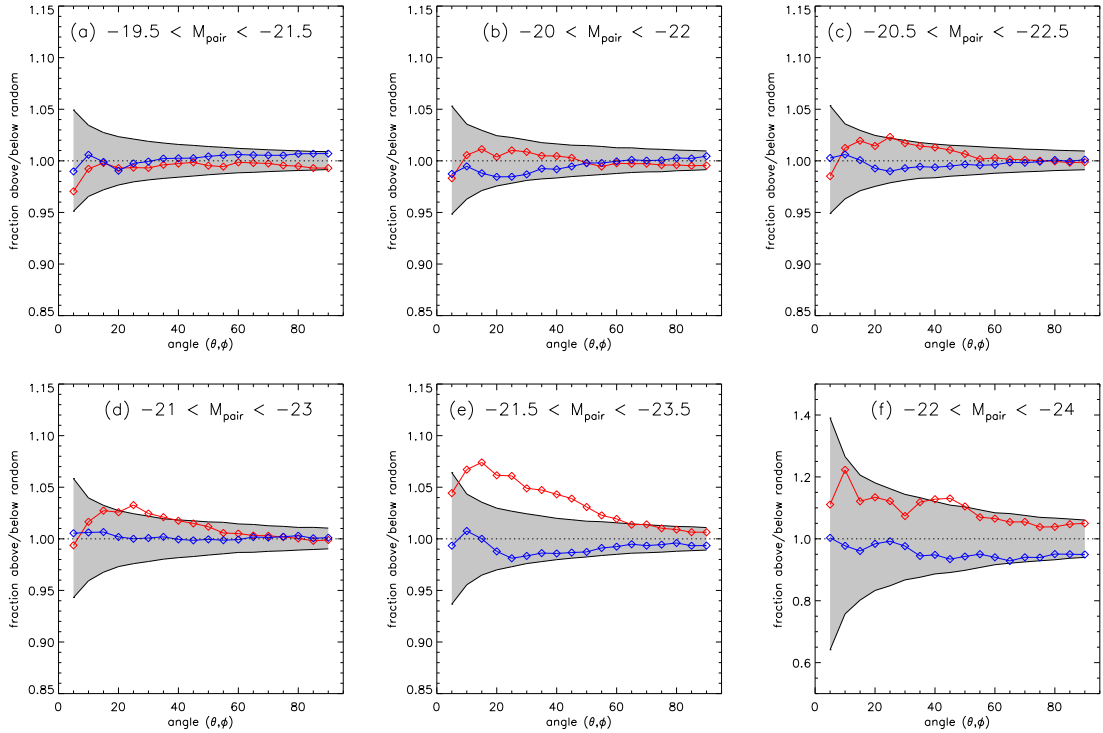}
   \caption{The signal is examined as a function of host magnitude. Pairs where both galaxy magnitudes fall within the range $-19.5<M_{\rm r}<-21.5$; $-20 <M_{\rm r}<-22$; $-20.5<M_{\rm r}<-22.5$; $-21<M_{\rm r}<-23$; $-21.5<M_{\rm r}<-23.5$ and $-22<M_{\rm r}<-24$ are shown in (a-f). The number of hosts that make the cut drops significantly in the highest magnitude bin (f). }
  \label{fig:magn}
 \end{figure*}

The reader will recall that we select only pairs that are within one magnitude of each other, but that fall in a bin two magnitudes wide such that our samples are not dominated by the low end of the luminosity function. In Fig.~\ref{fig:magn} we examine how the permitted brightness affects the observed signal, allowing it to vary in half magnitude increments from [$-19.5$, $-21.5$] to [$-22$, $-24$] in  Fig.~\ref{fig:magn}(\mbox{a--f}). Note that all other parameters ($r_{\rm search}$, $d_{\rm sep}$, etc.) are kept fixed at this stage and only the host magnitude is varied. 

There is no lopsided satellite distribution seen for lower magnitude pairs (Fig.~\ref{fig:magn}{a--c}), at least for the other parameters studied here. The signal starts to develop and slightly creep above the 3$\sigma$ significance for larger pairs (like the LG) in the bin [$-21$, $-23$], Fig.~\ref{fig:magn}(d) and is clearly strongest for the fiducial value [$-21.5$, $-23.5$], Fig.~\ref{fig:magn}(e). Since, few galaxies in the SDSS are brighter than $-23.5 < M_{r}^{0.0}$, changing the bright end limit has little effect on our results. As expected, the highest magnitude range, Fig.~\ref{fig:magn}(f), contains few galaxy pairs and thus the statistical significance is weak in this sample.

\subsubsection{Fore- and background galaxies as interlopers}
The lopsided signal measured in this work is a lower limit since it is undoubtedly contaminated by a myriad of fore and background galaxies. Indeed we have ignored the photometric redshift of the alleged ``satellites''. However if these are considered, galaxies whose photometric redshift is inconsistent with the spectroscopic redshift of the pair, can be safely eliminated as potential satellites. 

\begin{figure*}
 \centering
 \includegraphics[width=2.0\columnwidth]{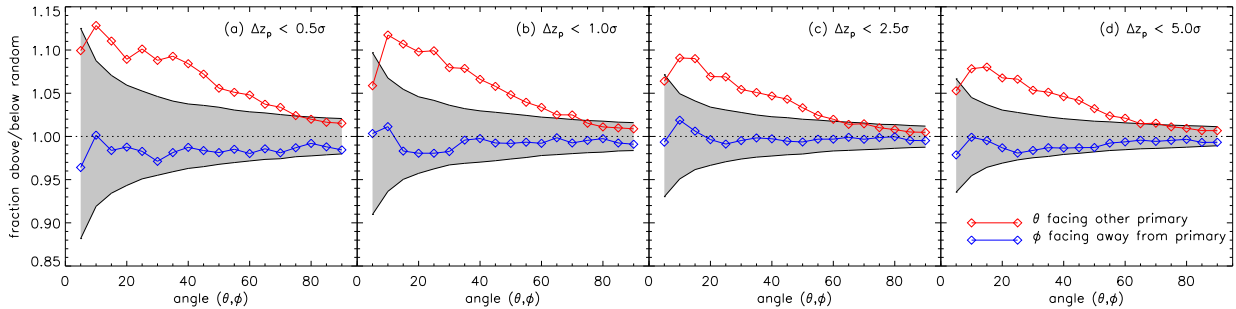}
   \caption{How photometric redshift cuts on the satellite sample affect the signal reported here. We attempt to assign satellite status to a galaxy based on whether the difference between it's photometric redshift and the pair's spectroscopic redshift, $\Delta z_{\rm p}$ is within 0.5$\sigma$ (a), 1.0$\sigma$ (b), 2.5$\sigma$ (c) or 5$\sigma$ (d) from the pair, where $\sigma$, is the error on the photometric redshift determination. The reader will note that the selection galaxies with $\Delta z_{\rm p} \lsim 2.5$ is generous enough that nearly all galaxies are included.}
  \label{fig:photoz}
 \end{figure*}

In Fig.~\ref{fig:photoz} we show how the lopsided signal depends on whether or not the photometric redshift of a satellite is consistent with the spectroscopic redshift of the pair. Specifically, in Fig.~\ref{fig:photoz}(a--d) we only consider a galaxy as a possible satellite if the spectroscopic redshift of the alleged host is within 0.5, 1, 2.5 or 5$\sigma$ (respectively) of the photometric redshift, where $\sigma$ is the error on the photometric redshift. When photometric redshifts are used to discriminate between interlopers and real satellites, the signal strength increases. Indeed when only galaxies whose redshift estimation is within 0.5$\sigma$ of the host, the signal is strongest. As expected as more and more interlopers are allowed to contaminate the signal, the signal weakens. When all galaxies whose photometric redshift is within 2.5$\sigma$ of the host, we essentially converge to the ``no photo-z cut'', since most interloping galaxies have redshifts consistent at this level. The reader will also note that when interlopers are excluded, the sample size is reduced and thus so is the statistical strength of any statement. In fact the significance of the signal in Fig.~\ref{fig:photoz}(a--d) is 4.4$\sigma$, 4.6$\sigma$, 4.2$\sigma$, and 4.4$\sigma$, respectively. Since the most conservative approach is to present a lower limit, we choose to preferentially present the case where the photometric redshift is ignored, noting that this is a lower limit.

\begin{figure*}
 \centering
 \includegraphics[width=2.0\columnwidth]{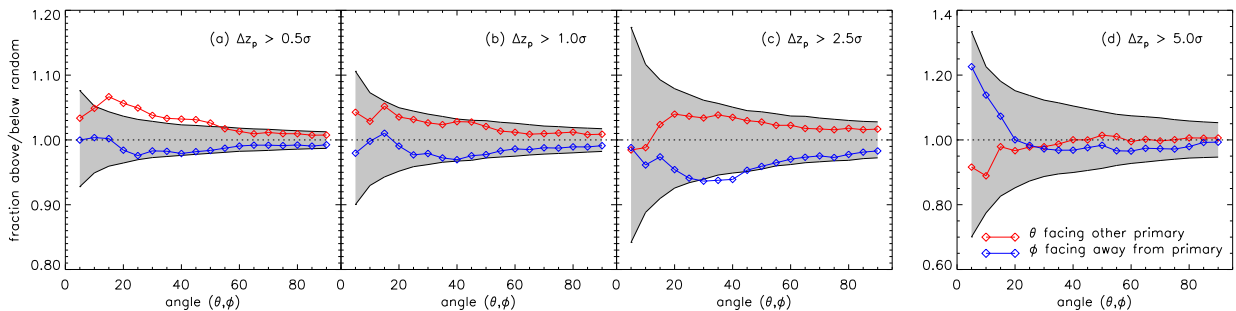}
   \caption{Similar to Fig.~\ref{fig:photoz}, here we show the uniformity of background objects, using photometric redshifts. As opposed to Fig.~\ref{fig:photoz}, where galaxies were {\it included} as satellites based on the difference between their photometric redshift and the pair's spectroscopic redshift ($\Delta z_{\rm p}$), here we {\it exclude} galaxies whose $\Delta z_{\rm p}$ is too large, specifically greater than 0.5$\sigma$ (a), 1$\sigma$ (b), 2.5$\sigma$ (c) and 5$\sigma$ (d). The reader will note that only when we (falsely) include galaxies with $\Delta z_{\rm p}<0.5\sigma$ does the ``background'' show any signal. In all other cases the, signal is perfectly consistent with the 3$\sigma$ spread one expects from uniform distributions. Note the difference in $y$-axis for this plot, compared with other plots in this paper.}
  \label{fig:interloper}
 \end{figure*}

The lopsidedness of just the background galaxies can also be qualified by examining galaxies whose photometric redshift implies that they are {\it inconsistent} with the spectroscopic redshift of the pair -- the opposite of Fig.~\ref{fig:photoz}. In Fig.~\ref{fig:interloper}(a--d), we examine the signal for galaxies whose photometric redshift is 0.5$\sigma$, 1$\sigma$, 2.5$\sigma$, and 5$\sigma$ (where $\sigma$ is the error of the photometric redshift) {\it greater} than the spectroscopic redshift of the pair, $\Delta {\rm z_{\rm p}}$. Since the $\Delta {\rm z}_{\rm p} > 0.5\sigma$ cut includes some satellites, in this regime the signal is still statistically significant ($\sim 2.3\sigma$). However for larger cuts the uniformity of the background becomes apparent as all the curves fall well within the $3\sigma$ spread expected for a uniform distribution.

\subsubsection{Search radius}
\label{sec:rsearch}
Perhaps the most interesting systematic we examine is the effect of the search radius $r_{\rm search}$ on the lopsided signal. In this case we remind the reader of the following methodology. A galaxy is first and foremost ascribed to the host to which it is closest. Namely, in the case one host has a satellite distribution which extends well into its partner's environs, such satellites will be assigned to the partner (i.e the overlap effect). Also, proximity in and of itself is not enough to be classified as satellite, one must also be within $r_{\rm search}$. Because galaxies are deemed to belong to those pair members to which they are closest, in the case of closely separated pairs or large $r_{\rm search}$ where $r_{\rm search}>d_{\rm sep}/2$, the search radius effectively becomes half the separation.

\begin{figure*}
 \centering
 \includegraphics[width=2.00\columnwidth]{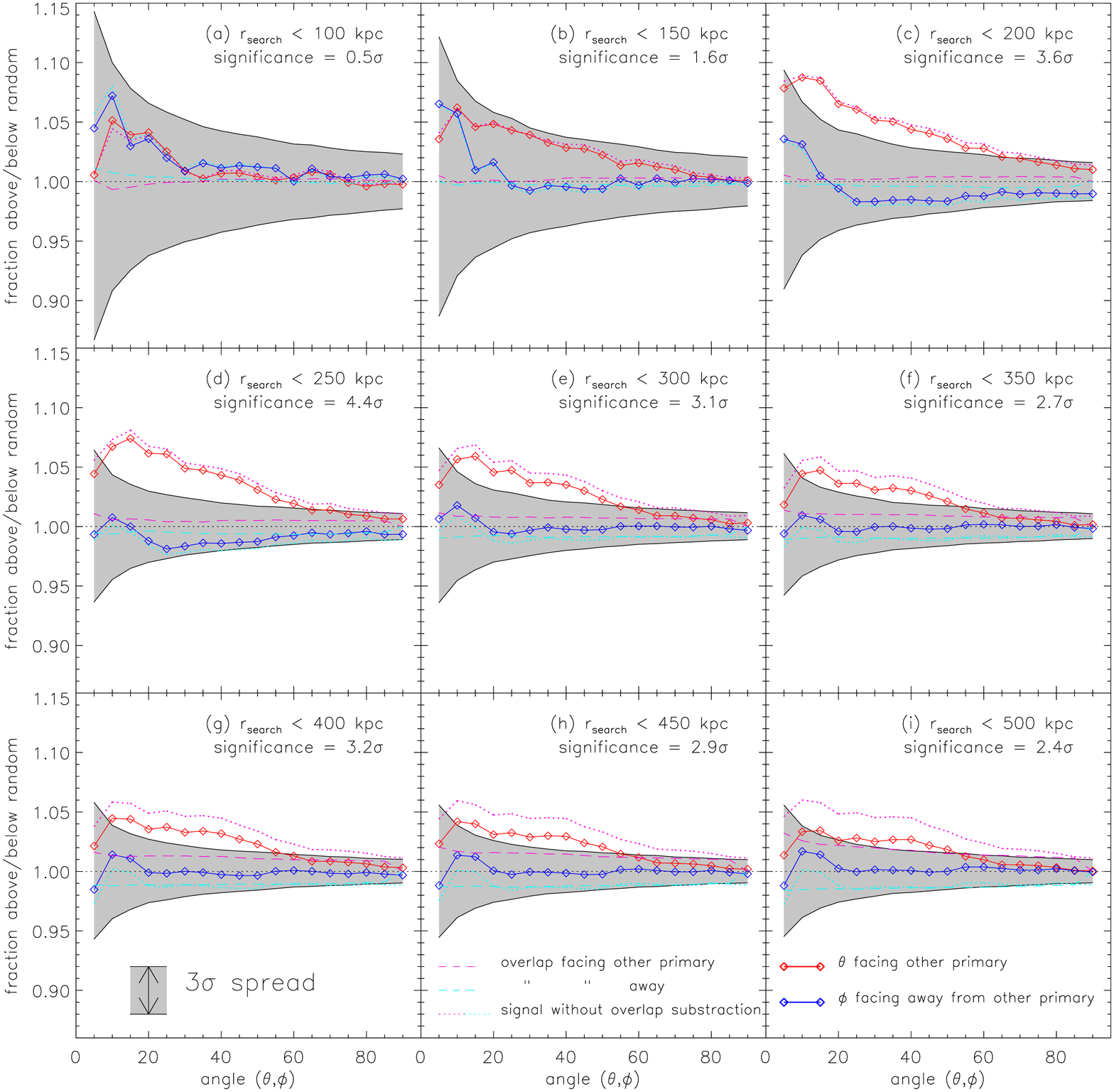}
   \caption{The signal varies as a function of the search radius $r_{\rm search}$ examined. For each satellite, the distance to each pair member is computed and it is assigned to the host to which it is closest. We show the signal for all satellites that fall within 100~kpc, 150~kpc, 200~kpc, 250~kpc, 300~kpc, 350~kpc, 400~kpc, 450~kpc, 500~kpc in (a)-(i) as the dotted magenta and cyan line (for the signal $\theta$ facing towards and $\phi$ facing away from the other pair member). The expected bias due to overlapping satellite distributions, computed by artificially placing isolated galaxies and their satellites at the same separation as the pair members, is shown by the dashed (cyan and magenta) lines. The full signal is computed as the measured signal (dotted lines) divided by the overlap bias (dashed lines) and is shown as the red and blue solid lines. The reader will note that as $r_{\rm search}$ increases so too does the overlap bias. The $3\sigma$ variance expected from a uniform distribution of the same size is shown as the grey shaded region. At each ($\theta,\phi$) we may compute the statistical significance of the signal and express it in units of $\sigma$, the 68\% variance. The average value of the significance is indicated in each panel.}
  \label{fig:rsearch}
 \end{figure*}

In Fig.~\ref{fig:rsearch}(a--i) we examine how the signal changes as $r_{\rm search}$ increases. We include the observed, overlap and true signal here so that the reader can see how all three of these vary with increasing $r_{\rm search}$. We highlight the salient points of this figure: both the observed and the overlap signal are effectively non-existent at $r_{\rm search}<100$~kpc and both increase significantly as $r_{\rm search}$ approaches 500~kpc, the largest distance allowed given the range of separations. This behavior is expected: in the closest parts around a host, the overlap effect is minimal as is any asphericity. The lopsided distribution of satellites gets stronger as $r_{\rm search}$ is increased, becoming visible (albeit statistically insignificant) already at 150~kpc. Statistical significance, defined as being greater than 3$\sigma$, emerges at 200~kpc and is maximum at 250~kpc (the fiducial value). The significance drops as $r_{\rm search}$ increases, but is still visible (yet statistically weak) at 500~kpc. Fig.~\ref{fig:rsearch}(a--i) implies that the signal is driven by anisotropy in both the 200--300~kpc range as well as within an opening angle of around 45$^{\circ}$.

\subsubsection{Pair separation}

\begin{figure*}
 \includegraphics[width=2.05\columnwidth]{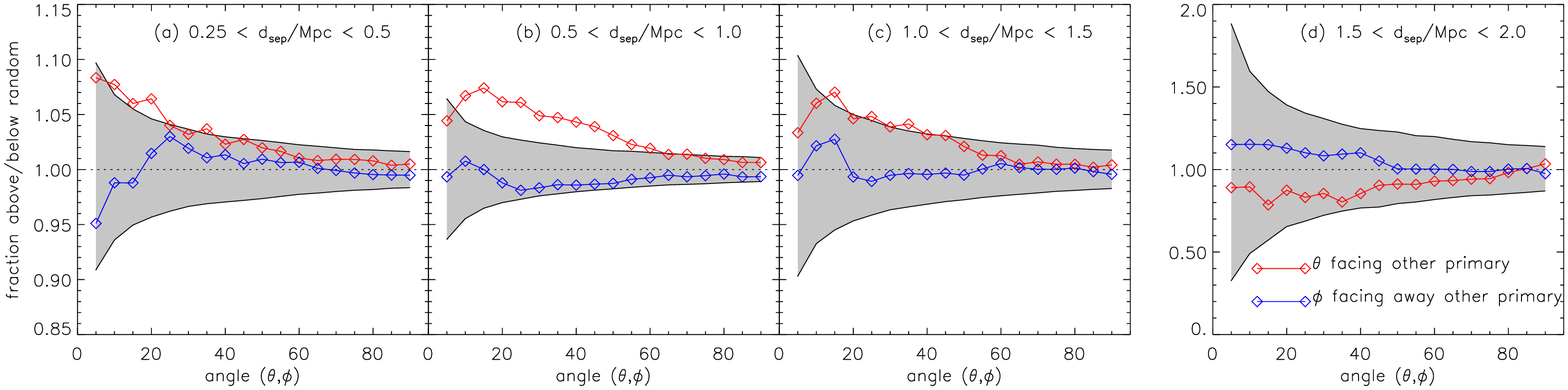}
  \caption{The signal is examined as a function of separation distance. Pairs where the hosts are separated by $0.25 {\rm Mpc}<d_{\rm sep}<0.5{\rm Mpc}$; $0.5{\rm Mpc}<d_{\rm sep}<1.0{\rm Mpc}$; $1.0{\rm Mpc}<d_{\rm sep}<1.5{\rm Mpc}$; and $1.5{\rm Mpc}<d_{\rm sep}<2.0{\rm Mpc}$;  are shown in (a-d). The number of hosts that make the cut drops significantly in the most widely separated bin (d). For the bin ($0.5{\rm Mpc}<d_{\rm sep}<1.0{\rm Mpc}$) used in our observed sample, the number of pairs and satellites is the largest.}
  \label{fig:dsep}
 \end{figure*}

The final criteria that could introduce a systematic effect is the pair separation $d_{\rm sep}$, chosen in the fiducial sample to be between 500 and 1000~kpc. Recall that in the above section~\ref{sec:rsearch}, when $r_{\rm search}$ is allowed to be large, we take as our effective search radius min($r_{\rm search},d_{\rm sep}/2$), a practice we follow here as $d_{\rm sep}$ is allowed to shrink. In Fig.~\ref{fig:dsep} we show the effect of changing the (projected) separation from closely separated pairs through to widely separated pairs. Signals are seen for pairs separated by [0.25, 0.5]~Mpc (\ref{fig:dsep}a), [0.5, 1.0]~Mpc (fiducial value, \ref{fig:dsep}b) and [1.0, 1.5] (\ref{fig:dsep}c), at the 2.2$\sigma$, 4.4$\sigma$ and $1.9\sigma$ levels respectively. Note that only few pairs are widely separated and make it into the final bin; here the sample size is small (a mere 440 satellites versus, e.g. 46\,043 satellites in the fiducial sample) and no signal is found. Although our results do not rule out lopsidedness being found in the slightly smaller or slighter larger separation bins, given this sample the statistical significance is less than the conventional 3$\sigma$ threshold. Perhaps given larger pair samples from future surveys, more statistically significant results may be obtained in these separation ranges.

\section{Summary and discussion}

Satellites of galaxies located in pairs identified in the SDSS show a statistically significant tendency to bulge towards one another. The identification of pairs is chosen to roughly mimic Local Group like systems namely to have two central galaxies of similar magnitudes and separation. All galaxies within 250~kpc (projected) of each primary -- a region similar to the virial radius -- are then chosen and the angle formed with the line connecting the two primaries is examined. In fact two angles are examined: the angle facing the other primary and the angle facing away from it. The number of galaxies within a given angle is then compared to that expected from a uniform distribution. 

Around $\sim8\%$ more galaxies are seen, within a $\sim 15^{\circ}$ angle facing the other primary than expected from a uniform distribution. Roughly the same number are seen in regions $180\degr$ away, namely in the direction opposite to the other primary. This discrepancy is the signature of a lopsided or bulging distribution. Monte-Carlo and Kolmogorov-Smirnov tests quantify the strength of the bulging signal as upwards of 4.4$\sigma$ -- simply put the chance that such a bulging is produced in random trials is less than one in 10 million. The lopsided distribution of satellites of pairs is thus a robust statistical signal.

Since galaxies can have extended satellite distributions, it is possible that this effect is being driven by a simple overlap of the two hosts' satellite populations. In order to distinguish a gravitational effect from an overlap, we conduct the exact same study on a purpose built ``overlap sample''. This is constructed by placing two isolated galaxies (chosen to have the same redshift and magnitude distribution as the main sample) and their satellites, at the same separation of each observed pair. The satellites of each of these isolated primaries are then randomized many times while keeping their radial distance from their host fixed, such that the effect of overlapping contamination can be gauged. Our overlap sample quantifies this overlap effect as being consistent with random at well below the 3$\sigma$ level. The overlap bias can be subtracted from the observed signal to reveal the ``true signal'', which is quantified by its average departure from uniformity, in units of $\sigma$, the standard deviation expected from random trials.

The possibility that some of the galaxy pairs studied here are simply isolated galaxies being observed along a cosmic filament is also considered. By using the SDSS filament catalog published by  \cite{2014MNRAS.438.3465T}, the nature of satellite distribution among pairs within and not in filaments was scrutinized. Filamentary environment is found to have a negative effect, namely the dynamically active filamentary environment destroys the lopsided signal. Similarly when pairs that are widely separated along the line of sight (namely in $\Delta z$) are selected, the lopsided effect weakens, hinting towards a gravitational origin.

In order to ascertain whether any of the specific choices that define the fiducial sample examined here were driving the lopsided signal, a systematic study was undertaken, wherein the lopsided effect was examined as a function of host magnitude and pair separation. In these cases, often the sub-sampling results in poor statistics and no credible signal. In the case where the search radius is increased, statistically viable lopsided signals are seen out to large radii.

Since our fiducial sample completely ignores the photometric redshift of the satellites, it is full of interlopers and the lopsided signal should thus be viewed as a lower limit. When the photometric redshifts are included, and satellites whose redshifts are clearly inconsistent with the host are excluded, the signal is significantly strengthened.

To some extent the alignment of satellites with the geometry of their parents is not wholly unexpected for two reasons. One is that the systems examined here are by construction members of galaxy pairs and thus not necessarily relaxed haloes in equilibrium, where spherical symmetry is assumed. Instead these may be merging, dynamically active pairs. Indeed the axial symmetry implicit in both the paradigmatic 1-dimensional NFW density profile \citep{1997ApJ...490..493N} as well as the extended triaxial  profile suggested by \cite{2002ApJ...574..538J}, assume different forms of isolation. It is unlikely that these conditions are met in the samples considered here (or in the Local Group for that matter). In these binary cases, we show that axial symmetry is clearly a poor assumption and that the amount of bulging -- at least in galaxy pairs -- can boost the number of satellites by up to around 10\%. 

The second reason why the position of satellite galaxies with respect to the geometry of their parents is not expected to be symmetric is related to the formation of haloes within the filamentary network known as the cosmic web. It is known that galaxy pairs and satellites are aligned with the filament they are embedded in \citep{2011MNRAS.411.1525L, 2011MNRAS.414.2029P, 2014ApJ...791...15L, 2015ApJ...799..212L, 2015MNRAS.452.1052L, 2015MNRAS.450.2727T, 2015ApJ...799...45F, 2015A&A...576L...5T, 2015MNRAS.450.2195S} and it is expected that these filaments channel satellites towards their hosts \citep{2014MNRAS.443.1274L, 2015MNRAS.452.1052L, 2004ApJ...603....7K,2016arXiv160106434G}. That said, these two empirical findings do not immediately and obviously imply axial asymmetry. In fact such a lopsidedness may be related to exchange of so-called ``renegade'' satellite galaxies among such pairs, as seen in simulations such as those of \cite{2011MNRAS.417L..56K}.

Our results indicate that triaxial modeling of the Milky Way's halo shape based, for example on the geometry of stellar streams \citep{2010ApJ...714..229L} is likely only an approximation in galaxy pairs such as the Local Group, as is mass modeling based on satellite proper motions (or Local Group timing arguments) since these model haloes as symmetric or axial-symmetric objects. Future studies will quantify how severe the lopsidedness in our own halo is and thus how inaccurate the halo shape and halo mass measurements are. 

Simulations of Local Group like pairs is the obvious place to look to see how such asymmetries are formed and how they evolve. Such a study would require analyzing a large cosmological volume (such that the number of galaxy pairs is large) with high resolution (in order to resolve substructures). It remains to be seen whether simulations can shed light on this effect, an avenue of investigation we are currently following.

The axial-asymmetry reported here is potentially related to anisotropic satellite galaxy planes. For one, the plane of satellites discovered in Andromeda \citep{2013Natur.493...62I} points to and is lopsided towards the Milky Way. It remains to be seen whether other planes of satellites show similar alignments. The fact that just 4 out of the brightest 11 Milky Way satellites point towards M31 could be indicative of a large Milky Way mass wherein its potential is deeper and therefore more resilient to tidal effects from M31.

In sum, our results show that for galaxy pairs the axial symmetric modeling of haloes may be incorrect. Such a complication undoubtedly limits the applicability of studies which ignore it, for example, in attempts to measure a halo mass or shape based on dynamical tracers. This is often applied to the Milky Way using her stellar streams or satellites. Secondly, we expect satellite planes in galaxy pairs not to be symmetric about their host but to be biased towards the other partner, as seen in M31. The binary nature of such systems can thus not be ignored when considering satellite distributions. The Milky Way and M31's satellite system should therefore not be viewed in isolation but must be considered as a single dynamical system.

\acknowledgments

The authors would like to thank and acknowledge Marcel Pawlowski and Rain Kipper for useful discussions. ET acknowledge the support by the Estonian Research Council grants IUT26-2, IUT40-2, and by the European Regional Development Fund (TK133).


%% copy editing.  Individual instruments can be provided in parentheses,
%% after the keyword, but they will not be verified.

%Facilities: \facility{Nickel}, \facility{HST(STIS)}, \facility{CXO(ASIS)}.



\bibliographystyle{apj} % style aa.bst
\bibliography{mybib} % your references Yourfile.bib




\end{document}